***O. S. MOSTOVYI***, PhD student, Department of Theory digital machine № 100, V.M. Glushkov Institute of Cybernetics of the NAS of Ukraine, Kyiv, Ukraine; e-mail: alex@mostovyi.net; ORCID: https://orcid.org/0009-0006-6687-866X

# CONTROL FLOW GRAPH RECOVERY FOR DYNAMICALLY LOADED CODE VIA SYMBOLIC LIBRARY RESOLUTION

Control Flow Graphs are one of the main data sources for software analysis that use dynamic and static software analysis methods. Protected software and modern malware increasingly depend on dynamic code loading techniques to evade static analysis. Usage of runtime dynamic linking mechanisms introduces unresolved indirect calls that stop static Control Flow Graph recovery. This serves to hide dynamic library that can be used for prevention of security analysis. To address this limitation, an analysis technique is proposed that combines symbolic execution with speculative library preloading to recover Control Flow Graphs from binaries by using dynamic loading. The methodology uses custom software hooks that intercept dynamic loading operations during symbolic execution and perform actual library loading into the analysis state. The module is based on a two-level architecture that stores interception functions and instruction tracking at the same time, all within a symbolic execution environment. To avoid executing potentially malicious code that dynamic instrumentation tools require, the analysis was conducted entirely through symbolic execution, making it safe for malware analysis. For evaluation a batch of 16 synthetic benchmarks was used, employing various obfuscation techniques including encrypted library names, network-triggered loading, environment-derived paths, multi-stage decryption chains, fileless execution and manual executable and linkable format parsing. The experiments results show that module recovers on average 29.8 % additional Control Flow Graph nodes and 26.5 % additional edges compared to static analysis alone, achieves 100 % precision and 100 % recall in library detection, with all discoveries validated through Frida-based dynamic instrumentation.

**Introduction.** Modern software systems, such as browser plugins or enterprise applications, generally use the dynamic code loading technique. The code that is loaded at runtime cannot be detected by static analysis, representing a fundamental problem in assessing program behavior. Recovering control flow through dynamically loaded code would provide significant improvement in the security analysis and verification of software systems. A Control Flow Graph (CFG) is a graph representation of all possible execution paths through a program [1]. The vertices of a CFG are basic blocks – contiguous sequences of instructions with a single-entry point and single exit point. Exit points are control transfer instructions: conditional branches, unconditional jumps, function calls, and returns. The edges represent possible successors of each basic block and are therefore directed. Fig. 1 illustrates a simple CFG.

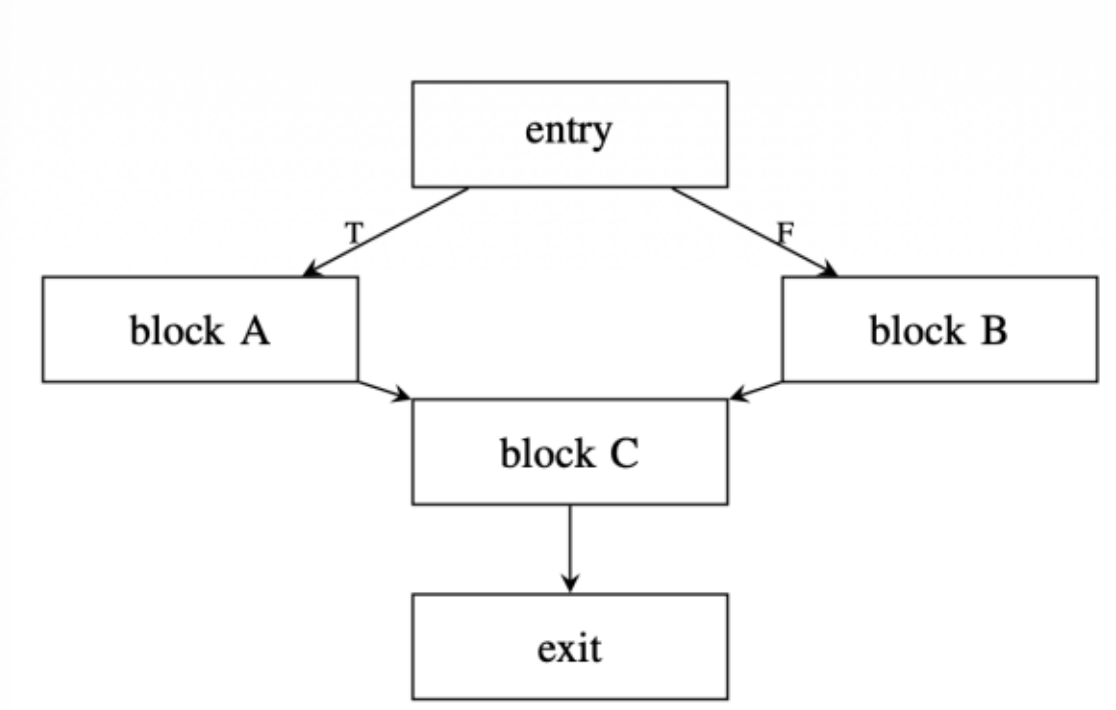

Fig. 1. A simple CFG with conditional branching.

Control Flow Integrity (CFI) is a security mechanism that checks a program's execution follows a predefined, valid path by using CFG [2,3]. This enables the detection and prevention attacks by ensuring that all indirect branches conform to legitimate edges in the CFG. The study [4] used CFG similarity to match functionally equivalent code across different architectures, which appears to be useful when searching for known vulnerabilities in binaries compiled for unfamiliar platforms. On the malware analysis side, CFG topology has become a common feature for classification, though as demonstrated in [5] demonstrated, models relying on these features can be fooled through adversarial manipulation. What these applications share is a common dependency: their reliability degrades when the underlying CFG is incomplete or imprecise, particularly in the presence of constructs like dynamic code loading. Programs load executable code through various mechanisms: high-level Application Programming Interfaces (API) invoke the dynamic linker to load shared libraries; system calls map files or anonymous memory with execute permissions; and manual loaders parse binary formats directly. Each mechanism can load code from diverse sources: filesystem, network, encrypted blobs, or memory making the loaded code invisible to static analysis. Dynamically loaded code executes through indirect control transfers whose targets depend on runtime values. A function pointer stores an address returned by symbol resolution; a virtual method dispatch indexes into a dynamically constructed table; a return instruction pops an address placed on the stack by loaded code. Static analysis sees only the indirect transfer instruction, not its concrete target. These levels interact: a library loading operation returns a handle; symbol resolution uses that handle to obtain a function address; an indirect call uses that address to transfer control. Analyzing either level in isolation fails as we must track both how code is loaded and where resolved addresses flow to indirect transfers. Existing analysis software resolves these levels separately. Static disassemblers (IDA Pro [6,7], Ghidra [8]) recover CFGs from existing code but cannot analyze code that will be loaded at runtime. Dynamic instrumentation tools (PIN [9], Frida [10]) observe runtime behavior but require executing the binary, which is dangerous for malware analysis and is limited to executed paths. Symbolic execution frameworks (Angr [1], S2E [11]) explore paths symbolically but treat dynamic loading APIs as symbolic values that lose all information about loaded code.

**Statement of the problem**. The primary challenge in program analysis is that dynamic code loading operates at two interacting levels, which existing tools analyze independently, resulting in incomplete CFGs. The aim of this study was to develop and assess a safe analysis module that recovers complete CFGs by simultaneously intercepting all code-loading APIs and monitoring indirect control flow at the instruction level via symbolic execution, correlating events across both levels to reconstruct the full chain from library loading through to indirect invocation.

**System-Level Evasion Techniques**. While traditionnal analysis assumes libraries are loaded via standard, documented POSIX interfaces [12], target programs actively manipulate lower-level operating system constructs to map executable code into memory. The most prevalent system-level evasion mechanisms include:

1. Internal API Abuse: Programs can bypass standard linker artifacts by directly invoking glibc-internal functions, such as __libc_dlopen_mode(). This allows library loading without importing dlopen into the Executable and Linkable Format (ELF) [13] dynamic section.
2. Fileless Execution via Anonymous Descriptors: Using the memfd_create system call, a program can allocate a RAM-backed file descriptor that does not exist on the physical filesystem. Writing binary data to this descriptor and executing it via /proc/self/fd/N completely bypasses any analysis that monitors at the filesystem.
3. Executable Memory Mapping: Programs can skip the dynamic linker altogether by utilizing open() to acquire a file descriptor, followed by mmap() or mmap64() with PROT_EXEC permissions.
4. Manual Relocation and Resolution: In the worst cases, a binary maps a shared library into memory and manually parses its ELF headers, maps individual segments, processes relocations, and resolves symbols independent of the OS loader.

Furthermore, the string parameters passed to these mechanisms exhibit varying degrees of resolution complexity. These range from simple hardcoded strings to highly complex paths that are decrypted at runtime, assembled from environment variables, or received dynamically over network sockets.

**Symbolic Execution**. Because the previous string parameters are often unresolvable through static disassembly, our method relies on symbolic execution. This approach replaces concrete inputs with symbolic variables, tracks path conditions as mathematical constraints, and utilizes Satisfiability Modulo Theories (SMT) solvers to explore feasible execution paths. Our implementation extends angr [1], a binary analysis framework. Our CFG construction relies on several core abstractions:

1. SimState: The fundamental unit of analysis, encapsulating the program's registers, file descriptors, and memory. Crucially, memory is modeled as a mapping to symbolic bitvectors, permitting simultaneous reasoning over both concrete addresses and symbolic data.
2. CLE (CLE Loads Everything): The framework's binary loader. CLE provides a dynamic_load() method that allows our module to inject additional libraries into the SimState execution, a critical requirement for simulating runtime loading.
3. SimProcedures and Claripy: SimProcedures are Python abstractions used to hook and replace specific library functions or system calls during execution. When these procedures encounter symbolic arguments, we use the Claripy constraint solving interface, built on Z3 [14], to concretize the values against a set of known candidate libraries.

**Architecture.** Our module is based on a two-level architecture that combines API-level interception and instruction-level tracking at the same time, all within a symbolic execution environment, as shown in Fig. 2. Each intercepted function is modeled as a SimProcedure subclass. The analysis proceeds in five stages:

1. The target binary is loaded into Angr.
2. 42 custom SimProcedures are installed at Procedure Linkage Table (PLT) stubs and known function addresses.
3. Symbolic execution explores the binary, triggering SimProcedures at API calls and VEX [15] breakpoints at indirect control transfers.
4. A correlation engine links events across both levels.
5. The output is a complete CFG covering both the main binary and all discovered libraries.

**Level 1 API**. It consists of 42 SimProcedures divided into six categories (Table 1). Each intercepted function is modeled as a SimProcedure subclass. We adopt the naming convention Dyn<FunctionName> for these classes. For example, DynDlopen replaces the default dlopen SimProcedure and extends it with library resolution logic.

Table 1 – Intercepted functions by category

| Category | Functions | Detection Purpose |
|---|---|---|
| Dynamic Loading | dlopen, dlsym, dlclose, dlmopen, dlvsym, dladdr, dlinfo | Standard library loading |
| Memory Ops | mmap, mmap64, mprotect, mremap, memfd_create | Executable mapping detection |
| Process Exec | execve, execveat, fexecve, clone, clone3 | Process replacement |
| Security | ptrace, prctl, setenv, putenv, process_vm_writev | Anti-debug, code injection |
| Network | socket, connect, recv, recvfrom, send, sendto, bind, listen, accept | Network payload interception |
| File Operations | open, openat, fopen, sigaction | File Directory Tracking |

The Dynamic Loading category (7 functions) handles the standard POSIX interface and its variants. DynDlopen extracts the path argument from dlopen(), resolves it through CLE's dynamic_load(), and returns a concrete handle address. DynDlsym looks up the requested symbol in the corresponding library and returns a concrete function address. DynDlclose decrements the handle reference count. We also intercept dlmopen for namespace-isolated loading, dlvsym for versioned symbol lookup, and the internal glibc function __libc_dlopen_mode. Concrete return values at this level are what allows Level 2 to resolve indirect call targets.

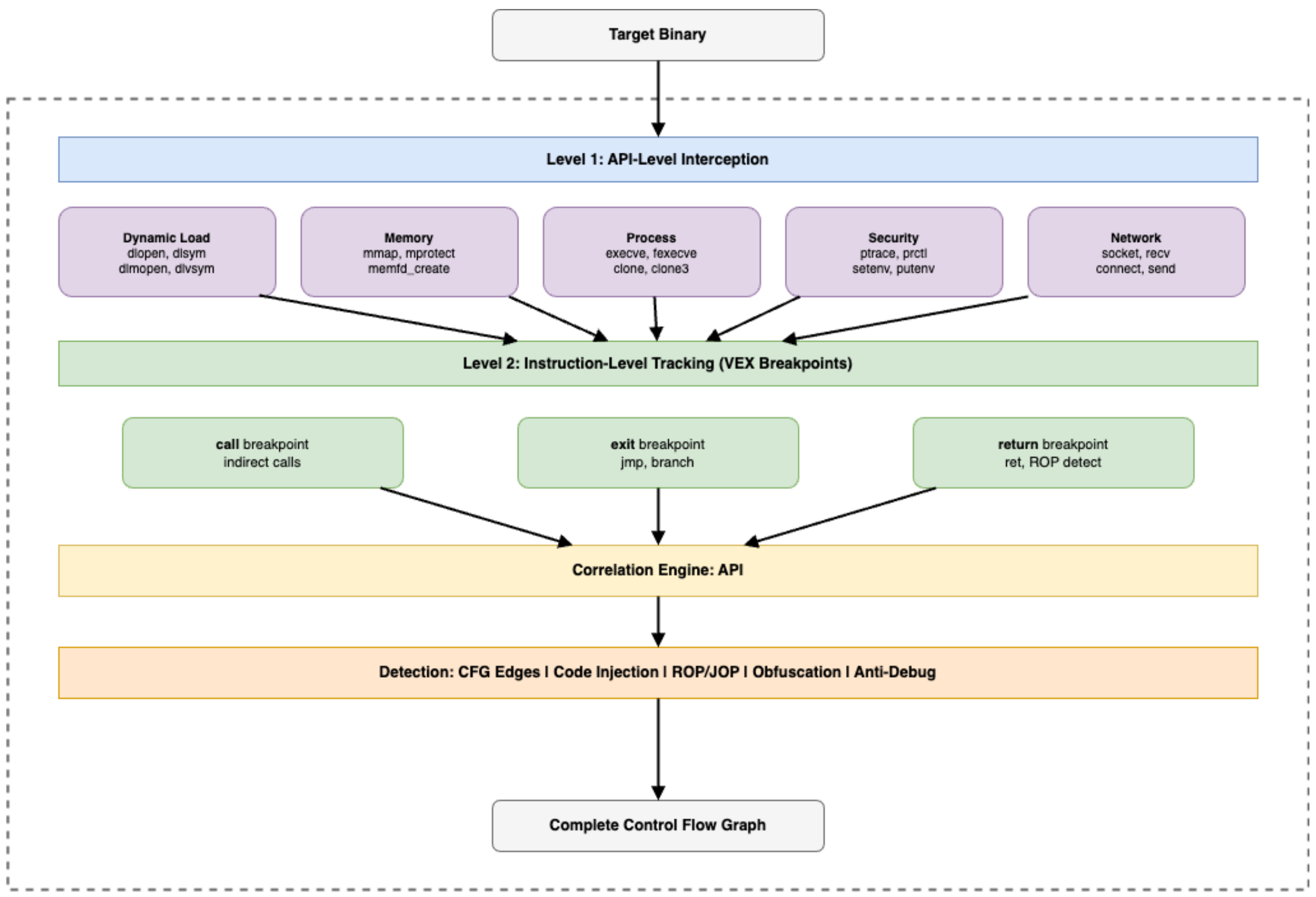

Fig. 2. Results of the population grouping

The Memory Operations category (5 functions) detects executable regions created outside the dynamic linker, open and openat record mappings between file descriptors and filesystem paths in an fd_to_path dictionary. When mmap or mmap64 is called with PROT_EXEC on a file-backed region, the corresponding path is retrieved from this dictionary. DynMprotect catches permission changes from writable to executable, DynMemfdCreate detects fileless loading, where a library is written to an anonymous memory-backed descriptor and loaded through dlopen(/proc/self/fd/N).

The Process Execution category (5 functions) covers execve, execveat, and fexecve for detecting process image replacement, as well as clone and clone3 for tracking new threads or processes that may load code independently.

The Security category (5 functions) intercepts ptrace to detect anti-debugging checks, prctl for process attribute changes, setenv and putenv for environment variable manipulation that may affect library search paths, and process_vm_writev for cross-process memory writes used in code injection.

The Network category (9 functions) intercepts socket, connect, recv, recvfrom, send, sendto, bind, listen, and accept. These are necessary for cases where the dlopen() argument is entirely received from a remote source and has no representation in the binary itself.

The File Operations category (4 functions) includes open, openat, and fopen for establishing file descriptors to file paths, providing data that the Memory Operations group needs. It also monitors sigaction for tracking signal handler registration that may trigger deferred library loading.

A central problem at this level is that path arguments passed to dlopen() may be symbolic, meaning they are encrypted, computed at runtime, or received from external input. The default angr SimProcedure for dlopen returns an unconstrained symbolic value and loses all loading information. Our replacement performs concrete resolution instead. All SimProcedures share the unified string extraction method presented in Fig. 3.

Algorithm 1 (UnifiedStringExtraction) handles three cases. If the pointer itself is symbolic, it returns a symbolic pointer indicator. If the loaded data is concrete, it decodes and returns the string directly. If the data is symbolic, the algorithm iterates over known candidate strings and checks whether each one satisfies the current path constraints through the SMT solver. When a satisfying candidate is found, the constraint is added and the string is returned. If no candidate matches, a fallback scans the binary for .so references, system library paths, and user search directories.

**Level 2 Instruction-Level Tracking**. It operates on the VEX intermediate representation and monitors indirect control flow transfers during symbolic execution [15]. Where Level 1 determines what is loaded, Level 2 determines where the resolved addresses are used. Before inter-pretation, angr lifts each machine instruction into VEX IR. Three breakpoint types registered on the symbolic state are:

**Algorithm 1** UnifiedStringExtraction

**Require:** State $s$, pointer $p$, max length $L$, encoding $E$
**Ensure:** Concrete string, symbolic AST, or error

1: **if** solver.symbolic($p$) **then**
2: **return** SYMBOLICPOINTER
3: **end if**
4: addr $\leftarrow$ solver.eval($p$)
5: data $\leftarrow$ memory.load(addr, $L$, $E$)
6: **if** $\neg$solver.symbolic(data) **then**
7: **return** decode(solver.eval(data), $E$)
8: **end if**
9: {Symbolic string: attempt speculative resolution}
10: candidates $\leftarrow$ GetPreloadedCandidates()
11: **for** each $c \in$ candidates **do**
12: **if** solver.satisfiable(data = encode($c$, $E$)) **then**
13: solver.add(data = encode($c$, $E$))
14: **return** $c$
15: **end if**
16: **end for**
17: **return** SYMBOLICSTRING(data)

Fig. 3. Algorithm 1 Unified String Extraction

1. Call breakpoints fire before every call instruction. If the target matches an address previously returned by module class DynDlsym or obtained through manual symbol lookup, a CFG edge is recorded from the call site to the target function.

2. Exit breakpoints fire before every jmp and conditional branch. They capture indirect jumps targeting dynamically resolved addresses and help identify control flow flattening dispatchers – blocks that branch to many successors based on a state variable.

3. Return breakpoints fire before every ret instruction. They examine the return address on the stack to detect Return-oriented programming (ROP) [16] redirection to dynamically loaded code.

When a breakpoint encounters a symbolic target, Algorithm 2 is applied, shown in Fig. 4. For each known executable region, the algorithm constrains the target to fall within the region's address range. If the constraint is satisfiable, the solver evaluates a concrete address under that constraint. The collected addresses form the set of possible call targets, all guaranteed to point to valid executable code.

**Algorithm 2** Symbolic Target Resolution

**Require:** State $s$, symbolic target $t$
**Ensure:** Set of concrete addresses

1: solutions $\leftarrow \emptyset$
2: regions $\leftarrow$ GETEXECUTABLEREGIONS()
3: **for** each $(start, end) \in$ regions **do**
4: $c \leftarrow (t \geq start) \wedge (t \leq end)$
5: **if** $s$.solver.satisfiable($c$) **then**
6: addr $\leftarrow s$.solver.eval($t$, extra = [$c$])
7: solutions $\leftarrow$ solutions $\cup$ {addr}
8: **end if**
9: **end for**
10: **return** solutions

Fig. 4. Algorithm 2 Symbolic Target Resolution

**Correlation Engine**. The correlation engine maintains three types of mappings across levels. First, file descriptors are mapped to filesystem paths – populated when Level 1 intercepts open calls and consumed when mmap is intercepted with PROT_EXEC. Second, library handles are mapped to loaded library objects – populated by class DynDlopen and consumed by other class DynDlsym. Third, resolved symbol addresses are mapped to their call sites – populated at Level 1 and matched at Level 2 to the indirect calls detected through breakpoints. The engine also performs taint tracking. Data received through recv() is marked as tainted, and if it later appears as a dlopen() argument, the engine records the complete data flow from network input to library loading.

Algorithm 3 (CFF Dispatcher Detection) searches the recovered CFG for control flow flattening patterns. Algorithm 4 (SMC and Inline Hook Detection) classifies write operations to executable memory. When such a write is detected , the original bytes are read and the new data is classified. A JMP and CALL pattern triggers a hook report with the computed relative target. A PUSH followed by RET – common in ROP chains – produces a PUSH or RET redirect report with the pushed address.

**Evaluation**. To enable complete evaluation, a suite of 16 benchmarks was created. These benchmarks cover a wide range of scenarios, from simple dynamic loading to advanced malware evasion techniques. For each benchmark, a dedicated binary was built to exercise a specific loading pattern. Table 2 shows which intercepted functions are involved in detecting each technique.

The module was assessed using a four-phase pipeline that compares a static baseline CFG with the CFG produced after dynamic library discovery (Fig. 5).

**Algorithm 5** Evaluation Pipeline

**Require:** Binary $b$, library search paths $\mathcal{P}$, ground truth $\mathcal{G}$
**Ensure:** Static CFG metrics, module CFG metrics, validation status

1:
2: **Phase 1: Static Baseline**
3: $P_s \leftarrow$ LOADBINARY($b$, auto_load_libs=False)
4: CFG$_s \leftarrow$ CFGFAST($P_s$)
5:
6: **Phase 2: Dynamic Library Discovery**
7: $P_m \leftarrow$ LOADBINARY($b$, auto_load_libs=False)
8: Install 42 SimProcedure hooks on $P_m$
9: $\sigma_0 \leftarrow$ ENTRYSTATE($P_m$)
10: simgr $\leftarrow$ SIMULATIONMANAGER($\sigma_0$)
11: **while** |simgr.active| $> 0$ **do**
12: simgr.STEP()
13: Prune active states to $\leq 32$
14: **end while**
15: $\mathcal{L} \leftarrow$ RECURSIVESCAN(DYNDLOPEN.LIBRARIES)
16:
17: **Phase 3: Module CFG Construction**
18: CFG$_m \leftarrow$ CFGFAST($P_m$) {now includes $\mathcal{L}$}
19:
20: **Phase 4: Validation**
21: **for all** $\ell \in \mathcal{L}$ **do**
22: FRIDAVALIDATE($b$, $\ell$)
23: **end for**

Fig. 5. Algorithm 5 Evaluation Pipeline

Both CFGs are generated through the same analysis – the only difference is that the module CFG includes libraries resolved and loaded by our module. The difference between the two graphs directly reflects the module's contribution. For each benchmark, recall is computed as the proportion of ground-truth libraries found, and precision as the proportion of discoveries that match the ground truth. The CFG improvement is measured through three metrics: additional nodes, additional edges, and additional functions. Every discovery is validated independently using Frida-based dynamic instrumentation.

**Environment**. All experiments are completed inside Docker containers running Ubuntu 22.04 LTS with Python 3.11 and Angr 9.2.98. The host system is an Apple MacBook with M-series ARM64 processor running macOS and Docker.

**Verification.** All 16 benchmark binaries are compiled as ELF 64-bit ARM64 executables using GCC 11.4 with default optimization level to preserve the original control flow structure of each evasion technique [17]. Payload libraries are compiled as independent shared objects. Each benchmark is executed in an isolated container instance to prevent issues between experiments. Analysis times are measured as durations averaged over three independent runs to account for variability in constraint solver performance. Table 2 presents the complete library detection results across all benchmarks together with independent Frida validation status. For each benchmark, the number of expected libraries is reported, the number of libraries discovered by our module, the Frida validation outcome, and the loading mechanism employed. Based on these results, the module achieves 100 % recall and 100 % precision across all benchmarks. All benchmarks discover libraries that are confirmed through Frida dynamic instrumentation, providing independent verification of the symbolic analysis results.

**Areas of Improvement**. While the results are encouraging, the module is far from complete, because our benchmarks are synthetic. They mirror documented malware techniques, but real obfuscated binaries are complex and less predictable and will require testing against samples from virus repositories. Our module can flag obfuscation but makes no attempt to undo it. Reversing these transformations to recover the original program logic is a harder problem, and it is unclear how far angr's symbolic execution can be pushed in that direction without fundamental changes to the engine. Platform coverage is another constraint. The current implementation handles Linux ELF binaries on x86-64 and ARM64 reasonably well, and we do not support other OS. Taint tracking also has known blind spots. The engine follows explicit data flows but ignores implicit flows through control dependencies, which means precision degrades when execution passes through hash functions, compression routines, or cryptographic operations. Whether that integration is practical, and whether analysts would adopt it, remains an open question we intend to explore.

**Conclusions**. The research presented in this paper demonstrates the feasibility of recovering control flow graphs from binaries that rely on dynamically loaded code whose identity is determined only at runtime. By leveraging a symbolic execution engine and extending it with a two-level interception architecture, our module can resolve library targets that are encrypted, constructed on the stack, derived from network sources, or loaded through fileless mechanisms, and subsequently construct CFGs that reflect the true program structure. The evaluation on a suite of 16 benchmarks shows that the module achieves 100 % recall and 100 % precision for library detection across all tested scenarios. The module CFGs produced after analysis contain, on average, 29.8 % more nodes, 26.5 % more edges, and 41.6 % more functions than the corresponding static baselines, demonstrating an improvement in CFG completeness. All discoveries are independently validated through Frida-based dynamic instrumentation. At the same time, the work highlights limitations and directions for further development. Obfuscation patterns are detected but not reversed, and taint propagation does not yet capture implicit information flows. Future iterations of the system could address these gaps by integrating automatic deobfuscation, extending platform support to additional OS, and incorporating machine learning classifiers for novel evasion technique identification. Overall, our study delivers a working demonstration of how symbolic execution can be combined with comprehensive API interception to recover

Table 2 – Verification Results

| Benchmark | Steps | Time, s | CFG Nodes | | CFG Edges | | Functions | | Loaded Objects | |
|---|---|---|---|---|---|---|---|---|---|---|
| | | | Static | Module | Static | Module | Static | Module | Static | Module |
| simple_dlopen | 20 | 0,21 | 67 | 101 | 73 | 106 | 37 | 57 | 5 | 7 |
| environment_path | 31 | 0,26 | 80 | 116 | 88 | 126 | 43 | 61 | 5 | 7 |
| xor_encrypted | 40 | 0,32 | 79 | 111 | 86 | 118 | 42 | 60 | 5 | 7 |
| computed_path | 33 | 0,26 | 84 | 115 | 93 | 124 | 44 | 61 | 5 | 7 |
| multi_stage | 20 | 0,25 | 77 | 151 | 91 | 164 | 37 | 81 | 5 | 9 |
| stack_strings | 22 | 0,25 | 72 | 101 | 78 | 109 | 39 | 54 | 5 | 7 |
| time_triggered | 33 | 0,29 | 75 | 104 | 85 | 116 | 39 | 54 | 5 | 7 |
| anti_debug | 48 | 0,32 | 112 | 141 | 130 | 161 | 53 | 68 | 5 | 7 |
| memfd_create | 26 | 0,34 | 136 | 165 | 158 | 189 | 67 | 82 | 5 | 7 |
| indirect_call | 36 | 0,26 | 73 | 102 | 88 | 119 | 34 | 49 | 5 | 7 |
| multi_encoding | 151 | 0,97 | 105 | 134 | 121 | 152 | 46 | 61 | 5 | 7 |
| manual_elf_load | 46 | 0,36 | 153 | 182 | 192 | 223 | 57 | 72 | 5 | 7 |
| mmap_exec | 49 | 0,37 | 119 | 152 | 144 | 177 | 50 | 69 | 5 | 7 |
| rop_chain | 71 | 0,40 | 99 | 131 | 118 | 151 | 47 | 65 | 5 | 7 |
| signal_handler | 71 | 0,41 | 114 | 176 | 138 | 202 | 53 | 87 | 5 | 9 |
| network_socket | 82 | 0,39 | 118 | 147 | 146 | 177 | 53 | 68 | 5 | 7 |

CFG edges that are invisible to static analysis. It not only validates the technical foundation of such an approach but also establishes a future research in binary analysis, malware detection, and software hardening can be built.

**Declaration on the use of generative AI.** During the preparation of this work, the authors used Grammarly for grammar and spell checking, as well as for rephrasing and reformulating the text. After using these tools, the authors reviewed and edited the content as necessary and take full responsibility for the content of this publication.

***О. С. МОСТОВИЙ***, аспірант, відділ теорії цифрових автоматів № 100, Інститут кібернетики імені В. М. Глушкова НАН України, Київ, Україна; e-mail: alex@mostovyi.net; ORCID: https://orcid.org/0009-0006-6687-866X

## ВІДНОВЛЕННЯ ГРАФІВ ПОТОКУ КЕРУВАННЯ ДЛЯ ДИНАМІЧНО ЗАВАНТАЖЕНОГО КОДУ ШЛЯХОМ СИМВОЛЬНОГО РОЗВ'ЯЗАННЯ БІБЛІОТЕК

Графи потоку керування є одним із ключових джерел даних як для статичного, так і для динамічного аналізу програмного забезпечення, оскільки саме вони відображають можливі шляхи виконання програми та слугують основою для виявлення вразливостей, верифікації поведінки й оцінки безпеки коду. Водночас захищене програмне забезпечення та сучасне шкідливе ПЗ дедалі активніше послуговуються механізмами динамічного завантаження коду й зв'язування під час виконання, що залишає невирішеними непрямі виклики та унеможливлює повне статичне відновлення графів потоку керування. Як наслідок, залежності від зовнішніх бібліотек залишаються прихованими, а результати статичного аналізу – неповними, що суттєво обмежує можливості аналізу безпеки бінарних файлів. Для подолання зазначеного обмеження запропоновано метод аналізу, що поєднує символічне виконання зі спекулятивним попереднім завантаженням бібліотек для відновлення графів потоку керування з бінарних файлів, які використовують динамічне завантаження. Методологія ґрунтується на застосуванні спеціальних програмних перехоплювачів, що фіксують операції динамічного завантаження під час символічного виконання та інтегрують фактично завантажені бібліотеки безпосередньо у стан аналізу, завдяки чому раніше невирішені непрямі виклики отримують конкретні цілі переходу. Модуль реалізовано на основі дворівневої архітектури, яка забезпечує одночасне функціонування механізмів перехоплення функцій та відстеження інструкцій у єдиному середовищі символічного виконання. Принципова відмінність від інструментів динамічної інструментації полягає в тому, що аналіз проводиться цілком у символічному середовищі без фактичного запуску потенційно шкідливого коду, що робить підхід безпечним для дослідження зловмисного програмного забезпечення. Оцінку виконано на 16 синтетичних тестах, що охоплюють різноманітні техніки обфускації: зашифровані назви бібліотек, завантаження ініційоване мережею, шляхи, отримані зі змінних оточення, багатоетапні ланцюжки дешифрування, безфайлове виконання та ручний розбір виконуваних файлів і форматів зв'язування. За результатами експериментів зафіксовано приріст у середньому 29,8 % вузлів та 26,5 % ребер графів потоку керування порівняно з виключно статичним аналізом, а також досягнуто 100 % точності та 100 % повноти у виявленні динамічно завантажуваних бібліотек, що підтверджено перехресною валідацією за допомогою динамічної інструментації на основі Frida.

*Повне ім'я автора / Author's full name*

Мостовий Олександр Сергійович / Mostovyi Oleksandr Serhiiovych